\documentclass[fleqn,10pt]{wlscirep}
\usepackage[utf8]{inputenc}
\usepackage[T1]{fontenc}
\usepackage{babel}
\usepackage{amsmath}
\usepackage{amssymb}
\usepackage{graphicx}
\usepackage{float}
\usepackage[symbol]{footmisc}
\usepackage[font=small,labelfont=bf,justification=justified,width=0.95\textwidth]{caption}

\newcommand{\un}[1]{\,\mathrm{#1}}

\title{Accurate thermal conductivities from optimally short molecular dynamics simulations}

\author[1]{Loris Ercole}
\affil[1]{SISSA -- Scuola Internazionale Superiore di Studi Avanzati, via Bonomea 265, 34136 Trieste, Italy}
\author[2]{Aris Marcolongo}
\affil[2]{Theory and Simulations of Materials (THEOS) and National Centre for Computational Design and Discovery of Novel Materials (MARVEL), \'Ecole Polytechnique F\'ed\'erale de Lausanne, 1015 Lausanne, Switzerland}
\author[1,*]{Stefano Baroni}
\affil[*]{baroni@sissa.it}

\begin{abstract}
The evaluation of transport coefficients in extended systems, such as thermal conductivity or shear viscosity, is known to require impractically long simulations, thus calling for a paradigm shift that would allow to deploy state-of-the-art quantum simulation methods. We introduce a new method to compute these coefficients from optimally short molecular dynamics simulations, based on the Green-Kubo theory of linear response and the cepstral analysis of time series. Information from the \emph{full} sample power spectrum of the relevant current for a \emph{single} and relatively short trajectory is leveraged to evaluate and optimally reduce the noise affecting its zero-frequency value, whose expectation is proportional to the corresponding conductivity. Our method is unbiased and consistent, in that both the resulting bias and statistical error can be made arbitrarily small in the long-time limit. A simple data-analysis protocol is proposed and validated with the calculation of thermal conductivities in the paradigmatic cases of elemental and molecular fluids (liquid Ar and H$_2$O) and of crystalline and glassy solids (MgO and a-SiO$_2$). We find that simulation times of one to a few hundred picoseconds are sufficient in these systems to achieve an accuracy of the order of $10\%$ on the estimated thermal conductivities.
\end{abstract}

\begin{document}
\flushbottom
\maketitle

\section{Introduction}
\label{sec:intro}
Our microscopic understanding of heat and mass transport in extended systems is rooted in the Green-Kubo (GK) theory of linear response,\cite{Green1954,Kubo1957} as applied to the Navier-Stokes equations for the densities of the conserved extensive variables,\cite{Kadanoff1963,Forster1975} which include energy, momentum, and the particle numbers for each molecular species. Transport coefficients are determined by the equilibrium fluctuations of the relevant currents and are in fact proportional to their autocorrelation times. Notwithstanding its beauty, rigor, and broad scope, the practical implementation of the GK theory is known to require very long molecular dynamics (MD) simulations, much longer in fact than the  current autocorrelation times one is required to evaluate.\cite{Schelling2002,Nevins2007,Jones2012,Zhang2015,Oliveira2017} Even though a number of expedients have been devised to cope with this problem,\cite{Jones2012,Oliveira2017,Chen2010} none of them is fully satisfactory in that in some cases they fail to provide rigorous criteria to estimate the accuracy resulting from a given MD trajectory, and in all cases the length of the required simulations is unaffordable with accurate but expensive quantum simulation techniques.\cite{Carbogno2017}

It has long been believed that the GK theory of heat transport could not be combined with modern simulation methods based on electronic-structure theory because energy densities and currents are inherently ill-defined at the quantum-mechanical level.\cite{Stackhouse2010b} This misconception has been recently challenged by a couple of papers showing that, while energy densities and currents are actually ill-defined classically no less than they are quantum-mechanically, the heat conductivity resulting from the GK formula is indeed well defined by virtue of a kind of \emph{gauge invariance} stemming from energy extensivity and conservation.\cite{Marcolongo2016,Ercole2016} This finding spurred a renewed interest in the quantum simulation of thermal conduction,\cite{Carbogno2017,Kang2017} and made it urgent to devise a data-analysis technique to make these simulations affordable, thus paving the way to the \emph{ab initio} simulation of heat transport.

In this paper we propose a data-analysis protocol leading to an asymptotically unbiased and consistent estimate of transport coefficients (\emph{i.e.} the bias and the statistical error can be made both arbitrarily small in the limit of long simulation times) and requiring shorter simulations than used so far. This protocol avoids any \emph{ad-hoc} fitting procedure and naturally provides an accurate estimate of the statistical error, thus lending itself to an easy implementation and automated use. While motivated by heat transport applications, our approach naturally applies to \emph{any} other transport properties that can be expressed, in a GK framework, in terms of time integrals of suitable autocorrelation functions, such as, \emph{e.g.}, ionic conductivities, viscosities, and tracer diffusivity, to name but a few. We test our procedure in the specific case of heat transport and we find that simulation times of one to a few hundred picoseconds are sufficient to achieve an accuracy of the order of $10\%$ on the estimated thermal conductivities in a wide range of systems at typical physical conditions. In Sec. \ref{sec:theory} we present the general theory of our approach; in Sec. \ref{sec:applications} we validate it with extensive benchmarks performed on the calculation of thermal conductivities in the paradigmatic cases of elemental and molecular fluids (liquid Ar and H$_2$O) and of crystalline and glassy solids (MgO and a-SiO$_2$). Sec. \ref{sec:conclusions} finally contains our conclusions.

\section{Theory}
\label{sec:theory}
Heat transport in insulators is determined by the dynamics of atoms, the electrons following adiabatically in their ground state, and the GK theory of linear response \cite{Green1954,Kubo1957} allows one to derive the thermal conductivity from an analysis of equilibrium, possibly \emph{ab initio}, MD trajectories. According to this theory, the thermal conductivity of a macroscopic and isotropic system is given by:

\begin{equation}
\kappa=\frac{1}{Vk_{B}T^{2}}\int_{0}^{\infty}\!\langle\hat{J}(t)\hat{J}(0)\rangle\, dt,\label{eq:Green-Kubo}
\end{equation}
where $V$ and $T$ are the system volume and temperature, $\hat{J}$ is one Cartesian component of the macroscopic heat flux,\cite{Hansen2013,Ercole2016} $k_{B}$ is the Boltzmann constant, and the brackets $\langle\cdot\rangle$ indicate ensemble averages over initial conditions. Here and in the following a caret, as in ``$\hat{J}$'', denotes a realization of a random variable, whereas properties of the underlying stochastic process, such as time correlation functions, are indicated without carets. According to Eq. (\ref{eq:Green-Kubo}), the heat conductivity is proportional to the zero-frequency component of the power spectrum (aka the power spectral density) of the heat flux: $\kappa=\frac{1}{2Vk_{B}T^{2}}S(f=0)$,

\begin{equation}
S(f)=\int_{-\infty}^{\infty}\! \mathrm{e}^{-i2\pi f t}G(t)\, dt,\label{eq:power-spectrum}
\end{equation}
where $G(t)=\langle\hat{J}(t)\hat{J}(0)\rangle$ is the equilibrium time-correlation function of the heat-flux process.

In practice, MD gives access to a discrete sample of the process, $\hat{J}_{n}=\hat{J}(n\epsilon)$, where $0 \leq n \leq N-1$, $\epsilon$ is the sampling period of the current, and $N$ the length of the time series, that we assume to be even. The Wiener-Khintchine theorem \cite{Wiener1930,Khintchine1934} states that, for large $N$, the power spectrum of the heat-flux process evaluated at $f_k=\frac{k}{N\epsilon}$ is proportional to the expected value of the squared modulus of the discrete Fourier transform of the corresponding time series:

\begin{align}
S(f_k)&=\langle \hat S_k \rangle, \label{eq:S-unbiased}\\
\hat{S}_{k}&=\frac{\epsilon}{N} \left |\hat{F}_{k} \right |^2, \label{eq:periodogram-def}\\
\hat{F}_{k}&=\sum_{n=0}^{N-1}\mathrm{e}^{-2\pi i\frac{kn}{N}}\hat{J}_n, \label{eq:Fk}
\end{align}
for $0 \leq k \leq N-1$. The \emph{sample spectrum} $\hat S_k$ is thus an estimator of the power spectrum of the process; in the statistics literature it is usually referred to as the \emph{periodogram}. The reality of the $\hat J$'s implies that $\hat F_k=\hat F^*_{N-k}$ and $\hat S_k=\hat S_{N-k}$, so that periodograms are usually reported for $0\leq k\leq \frac{N}{2}$ and their Fourier transforms evaluated as discrete cosine transforms.

The heat flux is an extensive quantity whose density autocorrelations are usually short-ranged. In the thermodynamic limit it can therefore be seen as the sum of (almost) independent identically distributed stochastic variables, so that, according to the central-limit theorem, its equilibrium distribution is Gaussian. A slight generalization of this argument allows us to conclude that the heat-flux process is Gaussian as well. The heat-flux time series is in fact a multivariate stochastic variable that, in the thermodynamic limit, results from the sum of (almost) independent variables, thus tending to a multivariate normal deviate. This implies that at equilibrium the real and imaginary parts of the $\hat F_k$'s defined in Eq. \eqref{eq:Fk} are zero-mean normal deviates that, in the large-$N$ limit, are uncorrelated among themselves and have variances proportional to the power spectrum evaluated at $f_k$. For $k=0$ or $k=\frac{N}{2}$, $\hat F_k$ is real and $\mathrm{var} (\hat F_k)=\frac{N}{\epsilon}S(f_k)$; for $k\notin\left\{ 0,\frac{N}{2}\right\}$, $\mathrm{var}(\mathfrak{Re}\hat{F}_k)= \mathrm{var}(\mathfrak{Im}\hat{F}_k) = \frac{N}{2\epsilon}S(f_k)$.  We conclude that in the large-$N$ limit the sample spectrum of the heat-flux time series reads:

\begin{equation}
\hat{S}_{k} = S \left(f_k\right)\hat{\xi}_{k}, \label{eq:periodogram-distribution}
\end{equation}
where the $\hat{\xi}$'s are independent random variables distributed as a $\chi_1^2$ variate for $k=0$ or $k=\frac{N}{2}$ and as one half a $\chi_2^2$ variate, otherwise. For the sake of simplicity, we make as though all the $\hat{\xi}_k$'s were identically distributed as one half a $\chi_2^2$ variate for all values of $k$, thus making an error of order $\mathcal{O}(1/N)$, which vanishes in the long-time limit that is being assumed in most of the arguments presented in this paper. In many cases of practical interest, multiple time series are available to estimate the power spectrum of a process. For instance, in equilibrium MD, a same trajectory delivers one independent time series per Cartesian component of the heat flux, all of which are obviously equivalent in isotropic systems. In these cases it is expedient to define a mean sample spectrum by averaging over $\ell$ different realizations:

\begin{equation}
{^{\ell}\hat{S}}_{k}=\frac{1}{\ell}\sum_{i=1}^{\ell}{\hat{S}}_{k}^{i}, \label{eq:mean-periodogram}
\end{equation}
where the index ``$i$'' enumerates the realizations. For $k \notin \{ 0,\frac{N}{2} \}$ the ordinates of the mean sample spectrum are therefore distributed as in Eq. \eqref{eq:periodogram-distribution}, with the $\hat\xi$'s being $\chi_{2\ell}^2$ variates, divided by the number of degrees of freedom:

\begin{equation}
^\ell{\hat\xi}_{k}\sim\frac{1}{2\ell}\chi_{2\ell}^{2}.\label{eq:chi-square-nu}
\end{equation}

Eq. (\ref{eq:periodogram-distribution}) brings both good and bad news. The good news is, the expectation of the periodogram of the time series is the power spectrum of the process, \emph{i.e.} the former is an unbiased estimator of the latter. The bad news is, this estimator is not consistent, \emph{i.e.} its variance does not vanish in the large-$N$ limit. This is so because a longer time series increases the number of discrete frequencies at which the power spectrum is sampled, rather than its accuracy at any one of them.  A consistent estimate of the zero-frequency value of the power spectrum (or the value at any other frequency, for that matter) can be obtained by segmenting a time series into several blocks of equal length and then averaging over the sample spectra computed for each of them. When the length of the trajectory grows large, so does the number of blocks, thus making the variance of the average arbitrarily small. In practice, the determination of the optimal block size is a unwieldy process that leads to an inefficient determination of the length of the trajectory needed to achieve a given overall accuracy. We adopt a different approach that allows us to obtain a consistent estimate of the zero-frequency value of the power spectrum from the statistical analysis of a \emph{single} trajectory sample (\emph{i.e.} no block analysis is needed) and such that the estimate of the trajectory length necessary to achieve a given accuracy is optimal.

Spectral density estimation from finite empirical time series is the subject of a vast literature in the statistical sciences, embracing both parametric and non-parametric methods.\cite{Stoica2005} In the following we propose a semi-parametric method to estimate the power spectrum of a stochastic process, based on a Fourier representation of the logarithm of its power spectrum (the ``log-spectrum''). The advantage of dealing with the log-spectrum, instead of with the power spectrum itself, is twofold. First and foremost, the noise affecting the former is additive, instead of multiplicative, thus making it simple and expedient to apply linear filters: limiting the number of components of the Fourier representation of the log-spectrum acts as a low-pass filter that systematically reduces the power of the noise and yields a consistent estimator of the log-spectrum at any given frequency. Second, as a bonus, the logarithm is usually smoother than its argument. Therefore, the Fourier representation of the logarithm of the power spectrum is more parsimonious than that of the spectrum itself.
 
Let  $^{\ell}\hat{L}_{k} =\log (^{\ell} \hat{S}_{k} )$ be the ``\emph{sample log-spectrum}'' of our time series. By taking the logarithm of Eq. \eqref{eq:periodogram-distribution}, $^{\ell}\hat{L}_{k}$ can be expressed as:

\begin{equation}
^\ell\hat{L}_{k} = \log\left(S(f_k) \right) + \log\left( ^\ell{\hat\xi}_k\right) = \log\left(S(f_k) \right) + \lambda_\ell + {^{\ell}\hat{\lambda}}_{k}, \label{eq:log-SPD}
\end{equation}
where 

\begin{equation}
\lambda_{\ell} = \int_0^\infty \log\left (\frac{\xi}{2\ell}\right ) P_{\chi^2_{2\ell}}(\xi) \, d\xi = \psi(\ell)-\log(\ell) \label{eq:lambda-ell}
\end{equation}
is the expected value of the logarithm of the ${^\ell}\hat\xi$ stochastic variables defined in Eq. \eqref{eq:chi-square-nu}, $P_{\chi^2_{2\ell}}$ is the probability density of a $\chi^2_{2\ell}$ variate, $^{\ell}\hat{\lambda}_k = \log\left( ^\ell{\hat\xi}_k\right)  - \lambda_\ell$ are zero-mean identically distributed independent stochastic variables, and $\psi(z)$ and is the digamma function.\cite{PolyGamma}
The variance of the ${^\ell}\hat\lambda$ variables is:

\begin{equation}
\sigma_{\ell}^{2}  = \int_0^\infty \log\left (\frac{\xi}{2\ell}\right )^2 P_{\chi^2_{2\ell}}(\xi) \, d\xi - \lambda_{\ell}^2 =\psi'(\ell),\label{eq:sigma2-ell}
\end{equation}
where $\psi'(z)$ is the tri-gamma function.\cite{PolyGamma} 

Eq. \eqref{eq:log-SPD} explicitly shows that the sample log-spectrum of a  time series is equal to the logarithm  of the power spectrum one wishes to evaluate (modulo a constant), \emph{plus} a (non-Gaussian) white noise. In order to eliminate the high-frequency components of the noise, and thus reduce its total power, we define the ``\emph{cepstrum}'' of the time series as the inverse Fourier transform of its sample log-spectrum:\cite{Childers1977}

\begin{align}
^{\ell} \hat C_{n} &= \frac{1}{N}\sum_{k=0}^{N-1} {^{\ell} \hat L_{k}}\mathrm{e}^{2\pi i\frac{kn}{N}}, 
 \label{eq:sample-cepstrum}
\end{align}
and its coefficients as the \emph{cepstral coefficients}. According to a generalized central-limit theorem  for Fourier transforms of stationary time series,\cite{Peligrad2010} in the large-$N$ limit, the (inverse) Fourier transform of the $^\ell{\hat {\lambda}}$'s, appearing in Eq. \eqref{eq:log-SPD} and implicitly in Eq. \eqref{eq:sample-cepstrum}, is a set of independent (almost) identically distributed zero-mean normal deviates. It follows that:

\begin{align}
^{\ell} \hat  C_{n} &= \lambda_{\ell} \delta_{n0} + C_{n} +  {^{\ell}\hat{\mu}}_{n},\label{eq:cepstrogram} \\
C_{n} &= \frac{1}{N}\sum_{k=0}^{N-1} \log\bigl (S(f_k) \bigr ) \mathrm{e}^{2\pi i\frac{kn}{N}}, \label{eq:C-nohat}
\end{align}
where $^{\ell}\hat{\mu}_{n}$ are independent zero-mean \emph{normal} deviates with variances $\left\langle ^{\ell}\hat{\mu}_{n}^{2}\right\rangle =\frac{1}{N}\sigma_{\ell}^{2}$ for $n\notin\left\{ 0,\frac{N}{2}\right\}$ and $\left\langle ^{\ell}\hat{\mu}_{n}^{2}\right\rangle =\frac{2}{N}\sigma_{\ell}^{2}$ otherwise. This result can be easily checked explicitly by using the definition of the discrete Fourier transform; the non-trivial extra information provided by the central-limit theorem is the asymptotic independence and normality of the $^{\ell}\hat{\mu}$'s. Similarly to the sample power spectrum, the cepstral coefficients are real, periodic, and even: $\hat C_{n} = \hat C_{N-n}$. If the log-spectrum, $\log\bigl (S(f_k) \bigr )$, is smooth enough, the number of non-negligible $C_n$ coefficients in Eq. \eqref{eq:C-nohat} is much smaller than $N$. Let us indicate with $P^*$ the smallest integer such that $C_n \approx 0$ for $P^* \le n \le N-P^*$, and let us restrict the discrete Fourier transform of the cepstrum defined in Eq. \eqref{eq:sample-cepstrum} to $n<P^*$ or $n>N-P^*$. The $k=0$ component of this restricted Fourier transform reads:

\begin{align}
 ^{\ell}\hat{L}_{0}^{*} &= {^\ell\hat{C}}_{0}+2\sum_{n=1}^{P^{*}-1}{^{\ell}\hat{C}}_{n}  \label{eq:L0*} \\
 &= \lambda_\ell+ \log(S_0) + {^{\ell}\hat{\mu}_{0}}+ 2 \sum_{n=1}^{P^*-1} {^{\ell}\hat{\mu}_{n}}.
\end{align}
We conclude that $^{\ell}\hat{L}_{0}^{*}$ is a normal deviate with expectation and variance

\begin{align}
\langle {^{\ell}\hat{L}_{0}^{*}}\rangle \equiv L_{\ell}^{*} &= \log(S_{0}) + \lambda_{\ell}, \label{eq:L*} \\
\sigma_\ell^{*}(P^{*},N)^{2} &=\sigma_{\ell}^{2}\frac{4P^{*}-2}{N}. \label{eq:sigma*}
\end{align}
Eqs. \eqref{eq:L*} and \eqref{eq:sigma*} are the main results of our method: they show how the logarithm of the transport coefficient can be estimated from the (inverse) Fourier coefficients of the sample log-spectrum, Eq. \eqref{eq:L*}, and how the resulting statistical error \emph{only} depends on the number of Fourier coefficients retained and the total simulation length, Eq. \eqref{eq:sigma*}.

The value of $P^{*}$ is a property of the stochastic process underlying the time series, and is therefore independent of $N$: for any given value of $P^{*}$ the variance $\sigma_{\ell}^{*2}$ tends to zero in the large-$N$ limit, and $^\ell\hat{L}^*_0-\lambda_{\ell}$ is thus a consistent estimator of $\log(S_{0})$. Notice that the absolute error on $L^*_{\ell}$ directly and nicely yields the relative error on $S_{0}$, which is proportional to the transport coefficient we are looking for. In general, all the cepstral coefficients are different from zero and assuming that many of them actually vanish introduces a bias. The optimal value of $P^{*}$ is the one for which the bias, which is a decreasing function of it, is of the order of the statistical error, which instead increases with $P^*$. Its choice is the subject of \emph{model selection} theory, another vast chapter in the statistical sciences.\cite{Claeskens2008} Among the several tests that have been devised to perform this task, we choose the optimization of the Akaike's information criterion (AIC),\cite{Claeskens2008,H.Akaike1974} as described below.

Given a model depending on $P$ parameters, $\theta = \{\theta_{1}, \theta_{2}, \cdots \theta_{P}\}$, the AIC is a sample statistic defined as

\begin{equation}
\mathrm{AIC}(P) =-2\max_{\theta}\log\mathcal{L}(\theta,P)+2P,\label{eq:AIC}
\end{equation}
where $\mathcal{L}(\theta,P)$ is the likelihood of the parameters. The optimal number of parameters is then determined as the argument of the AIC minimum:

\begin{equation}
P_A^* \equiv \arg\min_P\mathrm{AIC}(P) . \label{eq:P*}
\end{equation}
In the present case the parameters of the model are the $P$ coefficients $C=\{C_{0},C_{1},\cdots C_{P-1}\}$ as defined in Eq. \eqref{eq:C-nohat}, and the log-likelihood reads, up to additive terms independent of $P$ and $C$:

\begin{equation}
2\log\mathcal{L}(C,P) = -\frac{N}{2\sigma_\ell^{2}}\left(C_{0}+\lambda_{\ell}-\hat{C}_{0}\right)^{2}
 -\frac{N}{\sigma_\ell^{2}}\sum_{n=1}^{P-1}\left(C_{n}-\hat{C}_{n}\right)^{2} -\frac{N}{\sigma_\ell^{2}}\sum_{n=P}^\frac{N}{2} \hat{C}_{n}^{2}.
\end{equation}
Evidently, the above expression is maximized, for given $P$, by $C_{n}=\hat{C}_{n}-\delta_{n0}\lambda_{\ell}$ for $n=0,1,\cdots P-1$, and the corresponding value of the maximum is: $2\max_C\log\mathcal{L}(C,P)=-\frac{N}{\sigma_\ell^{2}}\sum_{n=P}^\frac{N}{2} \hat{C}_{n}^{2}$.
We conclude that the value of the AIC is:

\begin{equation}
\mathrm{AIC}(P)=\frac{N}{\sigma_\ell^{2}}\sum_{n=P}^\frac{N}{2} \hat{C}_{n}^{2}+2P. \label{eq:AIC-P}
\end{equation}
The value of $P$ that minimizes this expression is the optimal number of parameters in the Akaike sense, $P_A^*$, as defined in Eq. \eqref{eq:P*}.

The maximum frequency available for spectral/cepstral analysis is the Nyqvist frequency,\cite{Oppenheim1999} determined by the sampling period $\epsilon$ as $f_{\mathrm{Ny}}=\frac{1}{2\epsilon}$. Transport coefficients only depend on the low-frequency behavior of the spectrum, which is independent of $\epsilon$, as long as the latter is small enough as to avoid aliasing effects. For this reason it may prove convenient to eliminate the high-frequency portion of the spectrum ($f>f^*$) by applying a low-pass filter to the time series (\emph{e.g.} a moving average \cite{MovingAverage}) and then resample the latter with a sampling period $\epsilon^*=\frac{1}{2f^*}$, thus resulting in a time series of $N^*=N\frac{f^*}{f_{\mathrm{Ny}}}$ time steps.
The optimal number of cepstral coefficients resulting from Eqs. \eqref{eq:P*} and \eqref{eq:AIC-P}, as well as the error in the estimate of the transport coefficients resulting from Eq. \eqref{eq:sigma*}, depends in general on the choice of the cutoff frequency,  $f^*$. The smaller $f^*$, the smaller will presumably be the number of cepstral coefficients necessary to describe the log-spectrum to any given accuracy over such a shorter frequency range. However, the shorter length of the filtered time series, $N^*$, results in an increased variance of the estimator $^\ell{\hat L^*_0}$ defined in Eq. \eqref{eq:L0*}, according to Eq. \eqref{eq:sigma*}. Numerical experiments performed on the MD data reported in Sec. \ref{sec:applications} show that both the estimated value of $^\ell{\hat L^*_0}$ and its variance are actually fairly insensitive to the value chosen for the cutoff frequency, $f^*$, provided that the power spectrum of the re-sampled time series faithfully features the first band of the original spectrum (\emph{i.e.} the first prominent feature) and that this band is not too peaked at the origin.

Data analysis based on the theory presented above is straightforward. Given a current time series from a MD trajectory, the sample spectrum $^\ell\hat S_k$ is first determined from Eqs. \eqref{eq:periodogram-def}, \eqref{eq:Fk}, and \eqref{eq:mean-periodogram} 
and smoothed out using \emph{e.g.} a moving average \cite{MovingAverage} performed over a narrow frequency window, so as to reduce the statistical noise to a level where the shape of the power spectrum of the underlying process can be appreciated.
A cutoff frequency, $f^*$, is then determined so as to encompass the first band of the smoothed power spectrum. The cepstrum $^\ell\hat C_n$ is computed by Fourier analyzing the logarithm of the sample spectrum up the cutoff frequency thus estimated, as in Eq. \eqref{eq:sample-cepstrum}, and the optimal number of cepstral coefficients is determined using Eqs. \eqref{eq:P*} and \eqref{eq:AIC-P}. The logarithm of the zero-frequency value of the power spectrum is finally estimated from Eqs. \eqref{eq:L0*} and \eqref{eq:L*} as  $^\ell{\hat L}^*_0-\lambda_\ell$, and its variance from Eq. \eqref{eq:sigma*}.

Mind the difference between the moving average performed in the frequency domain to smooth out the power spectrum and that performed in the time domain, as suggested before, and acting as a low-pass filter. Spectral smoothing using a moving average in the frequency domain is common practice in the analysis of time series, and it actually provides a consistent estimator of the power spectrum. In fact, the number of frequencies falling within a window of given width increases linearly with the length of the series, so that the variance of the average decreases as the inverse of the product of the length of the series times the width of the window. The resulting spectral estimate is however biased by the variation of the signal within the window, thus strongly reducing the width of the windows that can be afforded. Moreover, both the bias and the statistical error are difficult to estimate, due to the multiplicative nature of the noise; therefore neither the plain periodogram nor a running average thereof are adequate for a quantitative estimate of the zero-frequency value of the power spectrum, which is proportional to the transport coefficient we are looking for.

\section{Applications and benchmarks}
\label{sec:applications}

In order to benchmark the methodology described above, we have applied it to the calculation of the thermal conductivity in four systems representative of different classes of materials, spanning from elemental and molecular fluids (Ar and $\mathrm{H_2O}$) to crystalline and glassy solids (MgO and a-SiO$_2$). Classical MD simulations were run using the LAMMPS package \cite{LAMMPS} with the setup described below. For liquid Ar we used a Lennard Jones potential as described in Ref. \citenum{Argon-FF} at a temperature $T\approx 220\un{K}$ and density $\rho=1.55\un{g/cm^3}$,  in a cubic supercell containing 864 atoms, with a time step $\Delta t=4\un{fs}$. For liquid H$_2$O we used a flexible model as in Ref. \citenum{Water-FF} at a temperature $T\approx 300\un{K}$ and density $\rho=1.0\un{g/cm^3}$, in a cubic supercell containing 180 molecules, with a time step $\Delta t=0.5\un{fs}$. For crystalline (FCC) MgO we used a Buckingham-plus-Coulomb potential as in Ref. \citenum{MgO-FF} at a temperature $T\approx 1000\un{K}$ and density $\rho=3.61\un{g/cm^3}$, in a $4\times 4\times 4$ simple cubic conventional supercell with 512 atoms, and a time step $\Delta t=0.3\un{fs}$. For  a-SiO$_2$ we used the potential proposed by van Beest, Kramer, and van Santen \cite{Silica-BKS-1990} as implemented by Shcheblanov \emph{et al.},\cite{Silica-BKS-2015} at a temperature $T\approx 1000\un{K}$ and density $\rho=2.29\un{g/cm^3}$,  in a supercell containing 216 atoms with a time step $\Delta t=1\un{fs}$. The glass model was obtained from a quench from the melt ($T\approx 6500\un{K}$) at a constant quenching rate of $5.5\times 10^{12}\un{K/s}$. Each system was equilibrated in the NVT ensemble at the target temperature for several hundred picoseconds; data were then collected in the NVE ensemble for trajectories whose length was chosen so as to represent realistic simulation runs that could be afforded using \emph{ab initio} MD ($\mathcal{T} = 100\un{ps}$ for Ar, H$_2$O, and a-SiO$_2$, and $\mathcal{T} =500\un{ps}$ for MgO). In order to compare our estimates of the transport coefficients and their statistical errors with reliable and statistically significant reference data, in all cases we ran much longer ($\approx 50\un{ns}$) simulations. This allowed us to compare our predicted conductivities with accurate values estimated from the direct  integration of the  time auto-correlation function in Eq. \eqref{eq:Green-Kubo}, as obtained from a block average \cite{Frenkel2001} performed over the long trajectory. In addition, we could collect abundant statistics of our estimator for the transport coefficients, Eq. \eqref{eq:L0*}, and validate its normal distribution specified by Eqs. \eqref{eq:L*}, \eqref{eq:sigma*}. 

\begin{figure}[!htb]
\centering
\includegraphics[scale=0.95]{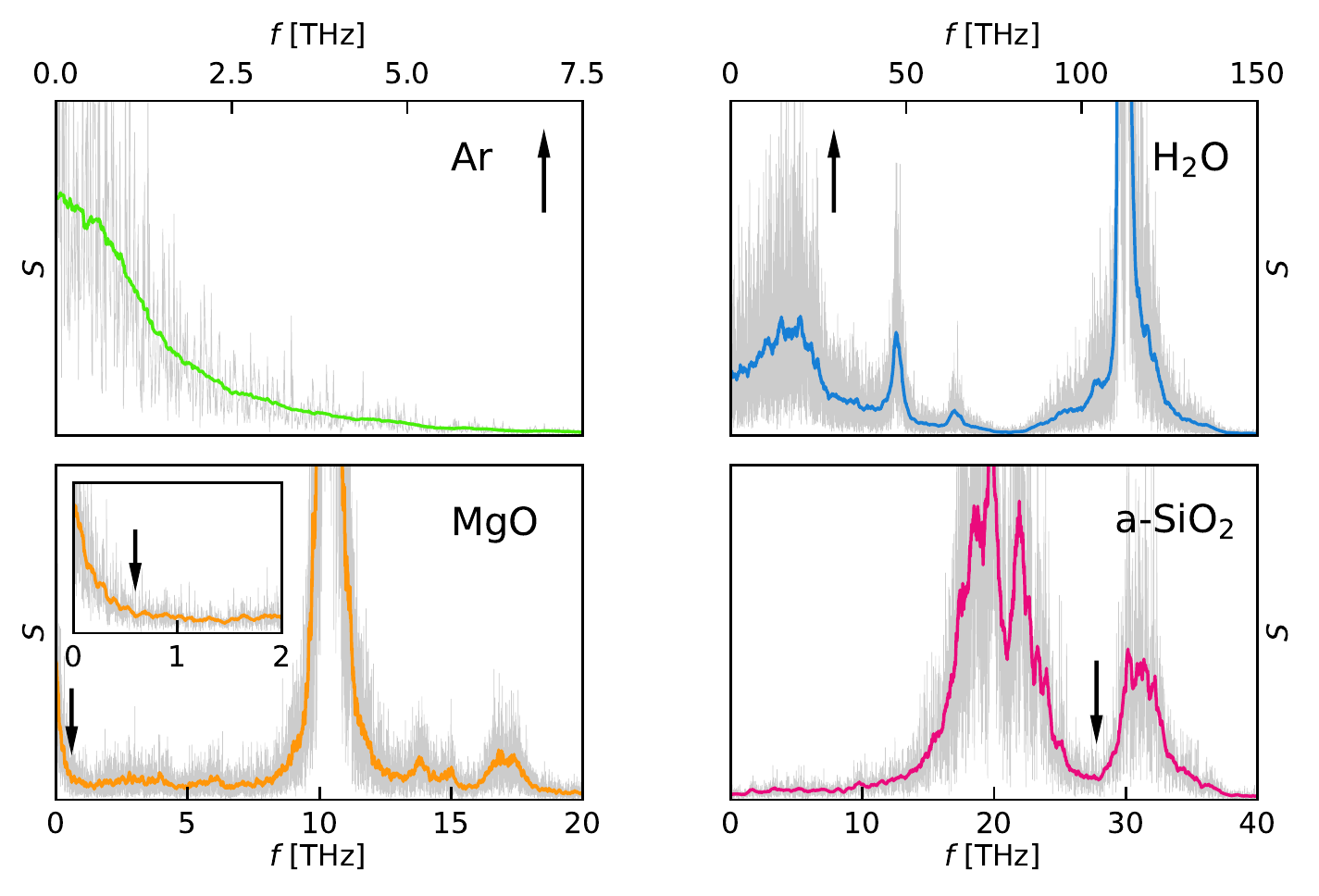}
\caption{Sample power spectra of the heat flux computed from MD trajectories for Ar, H$_2$O, a-SiO$_2$ ($100\un{ps}$) and MgO ($500\un{ps}$) (gray line, see text). The solid lines in color correspond to a moving average performed over a narrow frequency window. The vertical arrows indicate the cutoff frequencies, $f^*$, used for the subsequent cepstral analysis (see text). The inset in the MgO panel is a magnification of the low-frequency region of the spectrum.}
\label{fig:periodograms}
\end{figure}

In Fig. \ref{fig:periodograms} we report the sample periodograms of the heat fluxes computed for the four systems considered in this paper and averaged over the three Cartesian components, as described above. The solid lines in color indicate a further moving average \cite{MovingAverage} computed over a narrow frequency window, as explained at the end of Sec. \ref{sec:theory}. The values of the cutoff frequencies used for cepstral analysis, $f^*$, are chosen so as to encompass the first prominent feature of the (smoothed) power spectrum. For instance, in H$_2$O and a-SiO$_2$ we choose $f^*\approx 29\un{THz}$ and $f^*\approx 28\un{THz}$, respectively. In MgO we assume $f^*\approx 0.6\un{THz}$, just at the upper edge of the first narrow peak, whereas in Ar there is just one band, corresponding to a purely diffusive behavior of a simple fluid, and we assume $f^*\approx 7\un{THz}$, where the spectrum has exhausted most of the available power, but its value is not yet too small (see Fig. \ref{fig:periodograms}). The corresponding average numbers of cepstral coefficients given by the optimization of the AIC are: $P_A^*=5$ (Ar), $7$ (H$_2$O), $4$ (MgO), and $31$ (a-SiO$_2$). Later we will display the dependence of the number of optimal cepstral coefficients and of the resulting estimate of the thermal conductivity on the choice of $f^*$, and show that this choice is not critical.

\begin{figure}[!htb]
\centering
\includegraphics[scale=0.95]{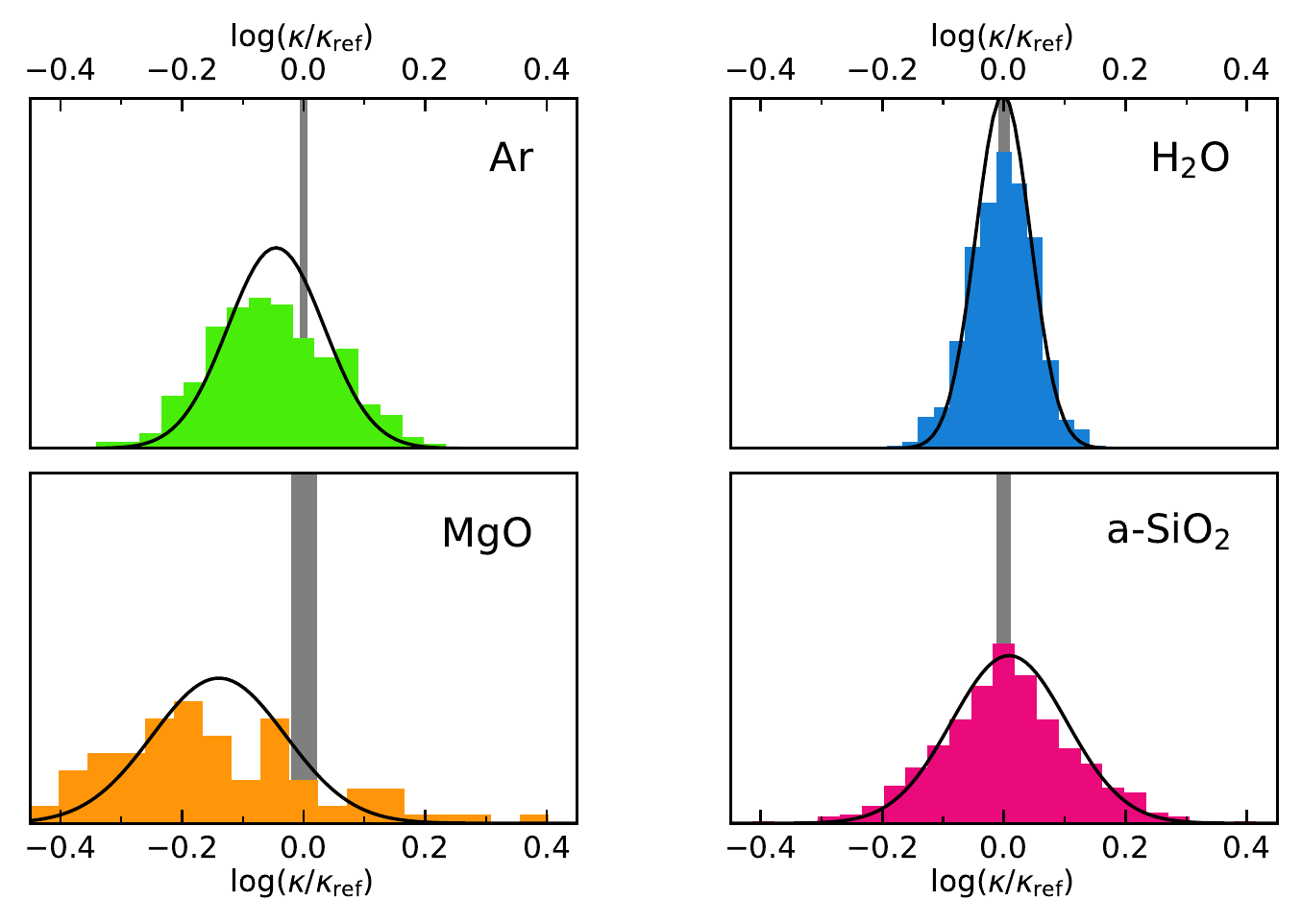}
\caption{Distributions of the logarithm of the thermal conductivities, $\log(\kappa)$, estimated over multiple MD segments ($100\un{ps}$ for Ar, H$_2$O, and a-SiO$_2$, and $500\un{ps}$ for MgO) extracted from a $50\un{ns}$ long trajectory. The reported data are referred to $\kappa_\mathrm{ref}$, which is the value obtained from the direct integration of the current autocorrelation function in Eq. \eqref{eq:Green-Kubo}, combined with standard block analysis over the $50\un{ns}$ trajectory, and represented by the vertical gray bands. The Gaussian curves represent the distributions predicted by the theory, centered at the sample mean. Remember that the absolute error on $\log(\kappa)$ is the relative error on $\kappa$.}
\label{fig:histograms}

\includegraphics[scale=0.95]{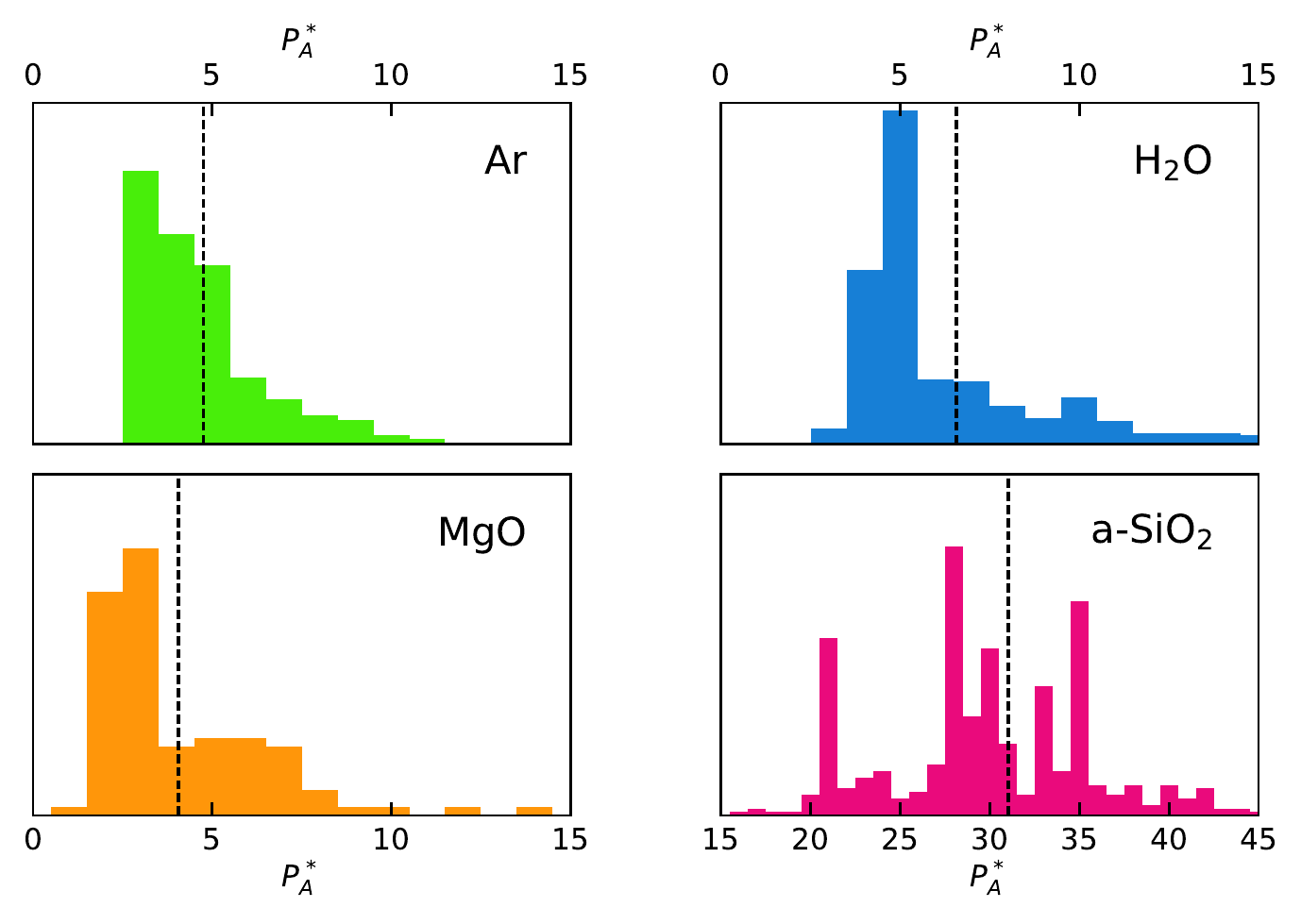}
\caption{Distribution of the optimal numbers of cepstral coefficients, $P_A^*$, as determined by optimizing the Akaike's information criterion, Eqs. \eqref{eq:P*} and \eqref{eq:AIC-P}, for each segment of the $50\un{ns}$ long MD trajectory, as described in the text. The vertical dashed lines indicate the average value of $P_A^*$.}
\label{fig:Pstar_distribution}
\end{figure}

\begin{figure}[!ht]
\centering
\includegraphics[scale=0.95]{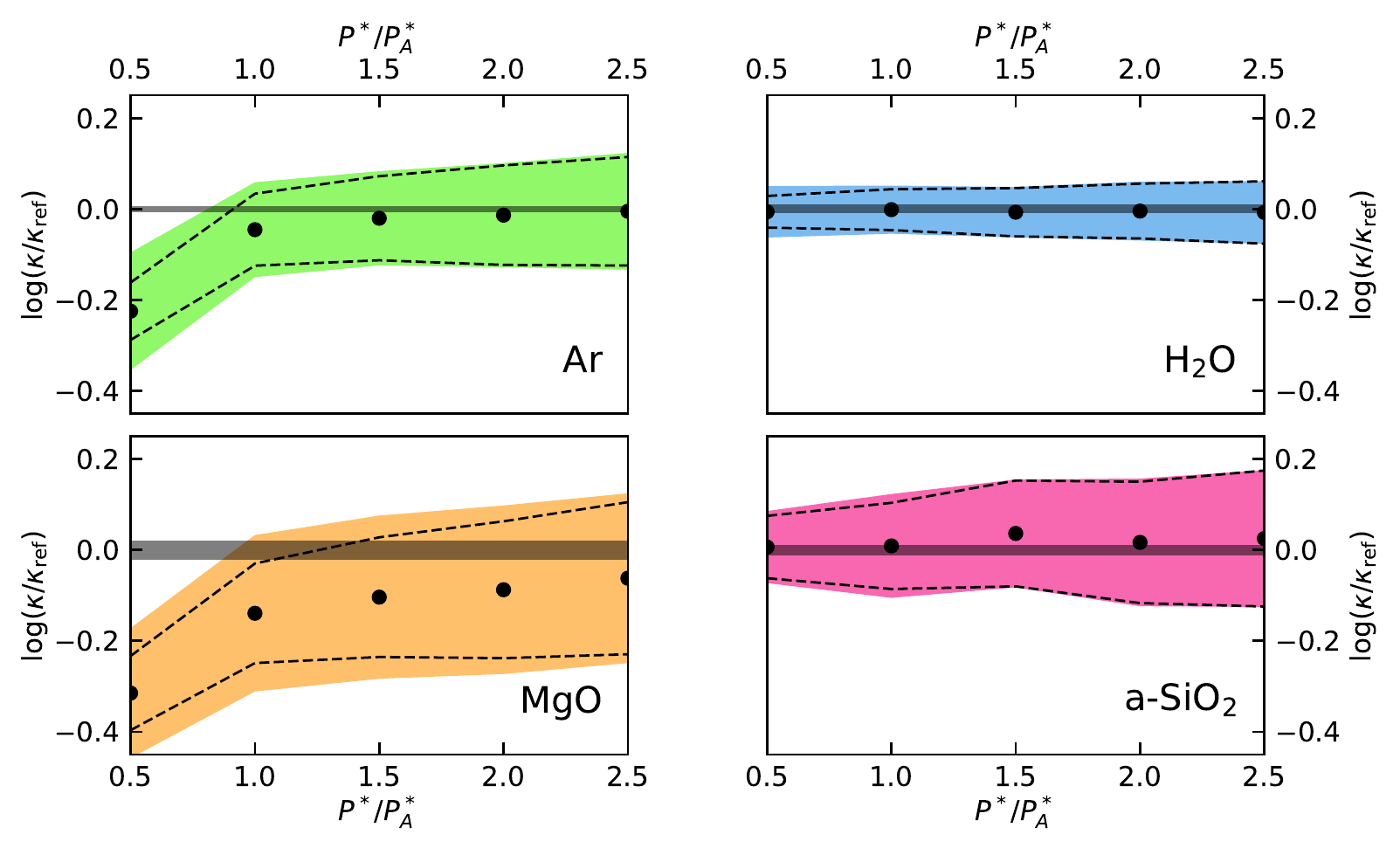}
\caption{Dependence of $\log(\kappa)$, as estimated from Eq. \eqref{eq:L0*}, on the number of cepstral coefficients $P^*$. $P_A^*$ is the optimal number of coefficients estimated from the AIC using Eqs. \eqref{eq:P*} and \eqref{eq:AIC-P}. The black dots represent the mean values of $\log(\kappa)$ computed over multiple MD segments ($100\un{ps}$ for Ar, H$_2$O, and a-SiO$_2$, and $500\un{ps}$ for MgO) extracted from a $50\un{ns}$ long trajectory; the colored bands and dashed lines represent one standard deviation as estimated from the empirical statistics and from Eq. \eqref{eq:sigma*}, respectively. The reported data are referred to $\kappa_{\mathrm{ref}}$, which is the value of thermal conductivity obtained from direct integration of the current autocorrelation function in Eq. \eqref{eq:Green-Kubo}, combined with standard block analysis over the $50\un{ns}$ trajectory, and represented by the horizontal gray bands. Remember that the absolute error on $\log(\kappa)$ is the relative error on $\kappa$.}
\label{fig:L-vs-P}
\end{figure}

In order to validate our data-analysis protocol, we first computed the heat conductivities from a direct integration of the current autocorrelation function, Eq. \eqref{eq:Green-Kubo}, combined with standard block analysis over the $50\un{ns}$ long trajectory, that will be taken as a reference, obtaining: $\kappa_{\mathrm{ref}} = 0.1965 \pm 0.0015$, $0.970 \pm 0.009$, $19.2 \pm 0.4$, and $2.115 \pm 0.025 \un{W/mK}$, for Ar, H$_2$O, MgO, and a-SiO$_2$, respectively. 
Although our simulations were meant for benchmarking purposes only, and no particular care was paid to exactly match the simulation conditions of previous work, these data are in fair agreement with the foregoing theoretical results: $\approx 0.19\un{W/mK}$ (Ar),\cite{Argon-FF} $\approx 0.85\un{W/mK}$ (H$_2$O),\cite{Romer2012} $\approx 12\un{W/mK}$ (MgO),\cite{MgO-FF} and $\approx 2.1\un{W/mK}$ (a-SiO$_2$).\cite{Larkin2014}

In Fig. \ref{fig:histograms} we display the distributions of the values of $\log(\kappa/\kappa_{\mathrm{ref}})$ estimated by applying our protocol to multiple MD segments of $100\un{ps}$ (for Ar, H$_2$O, a-SiO$_2$) and $500\un{ps}$ (for MgO), extracted from the $50\un{ns}$ long trajectory. The optimal numbers of cepstral coefficients, $P^*$, have been redetermined for each segment independently, while the values of the cutoff frequency, $f^*$, which only depends on the qualitative features of the spectrum, have been determined once for all for one of them. The distribution of the resulting number of cepstral coefficients is reported in Fig. \ref{fig:Pstar_distribution}. The observed distributions of $\log(\kappa)$ successfully pass the Shapiro-Wilk normality test \cite{Shapiro1965} (\emph{i.e.} they do not fail it) and the observed sample standard deviations closely match the theoretical values estimated from Eq. \eqref{eq:sigma*} (sample/theory):  $0.104/0.079$ (Ar), $0.053/0.045$ (H$_2$O), $0.17/0.11$ (MgO), and $0.114/0.095$ (a-SiO$_2$). Remember that the error on $\log(\kappa)$ is the relative error on $\kappa$: the corresponding absolute errors achievable with a short trajectory of $100$ (Ar, H$_2$O, and a-SiO$_2$) or $500$ (MgO) ps, not to be confused with the long trajectory used to establish the reference data above, are therefore: $\sigma_\kappa \approx 0.015$ (Ar), $0.045$ (H$_2$O), $2$ (MgO), and $0.2$ (a-SiO$_2$) $\un{W/mK}$. This indicates that a \emph{single} and short sample trajectory, such as one that is affordable with \emph{ab initio} MD, is sufficient to achieve and accurately estimate a very decent relative error on the computed transport coefficient. In an attempt to evaluate the thermal conductivity from the direct computation of the GK integral, Eq. \eqref{eq:Green-Kubo}, and standard block analysis using similarly short MD trajectories, our best estimate of the resulting statistical error was 2-3 times larger than using our protocol (meaning $5-10\times$ longer trajectories to achieve a comparable accuracy) in all cases but liquid Ar, where only a marginal improvement is achieved using our methodology. Much more than this, the standard analysis of MD data depends on a number of hidden parameters, such as the upper limit of the GK integral, Eq. \eqref{eq:Green-Kubo}, or the width of the blocks for error analysis, that are hard to determine and keep under control. Our method, instead, only depends on a single parameter, the number of cepstral coefficients, whose optimal value can be easily determined from the Akaike's information criterion, or other more sophisticated model-selection methods,\cite{Burnham2004,Claeskens2008} as appropriate.

In order to estimate the bias introduced by limiting the number of cepstral coefficients, we examined the sample mean computed over the distributions displayed in Fig. \ref{fig:histograms}, $\langle\kappa\rangle$, obtaining: $\langle\kappa\rangle = 0.1878 \pm 0.0007$, $0.969 \pm 0.002$, $16.7 \pm 0.2$, and $2.131 \pm 0.009 \un{W/mK}$, for Ar, H$_2$O, MgO, and a-SiO$_2$, respectively. Comparing these data with the reference data obtained from  the direct evaluation of the GK integral, we see that the bias is negligible for H$_2$O and a-SiO$_2$, very small for Ar, and small but not negligible for MgO. In Fig. \ref{fig:L-vs-P} we display the dependence of $\log(\kappa/\kappa_{\mathrm{ref}})$ on the number of cepstral coefficients, $P^*$, as estimated from Eq. \eqref{eq:L0*}. We observe that when the number of cepstral coefficients, $P^*$, is larger than the optimal value determined from the AIC, $P_A^*$, the estimated value of $\kappa$ seems not to depend on $P^*$ for all systems but MgO, for which a slight bias seems to persist, and to a much lesser extent for Ar. Also, Eq. \eqref{eq:sigma*} seems to slightly underestimate the sample variance for small $P^*$ in these cases. In the case of MgO this behavior is likely due to the difficulty of the AIC to cope with the sharp low-frequency peak in the power spectrum, due to the highly harmonic character and slow decay of the vibrational heat carriers in periodic crystals,\cite{Carbogno2017} thus requiring longer simulation times. In the case of Ar the very small bias observed for $P^* =P^*_A$ may be due to the difficulty of choosing a suitable cutoff frequency when only a single diffusive band is present in the spectrum, and to the divergence of the log-spectrum at high frequency. In all cases, use of the Aikake's information criterion results in a bias that is smaller than the statistical error estimated from an individual short sample trajectory and that can be systematically removed by increasing the value of $P^*$, at the price of increasing the statistical error, if and when needed.

\begin{figure}[!htb]
\centering
\includegraphics[scale=0.95]{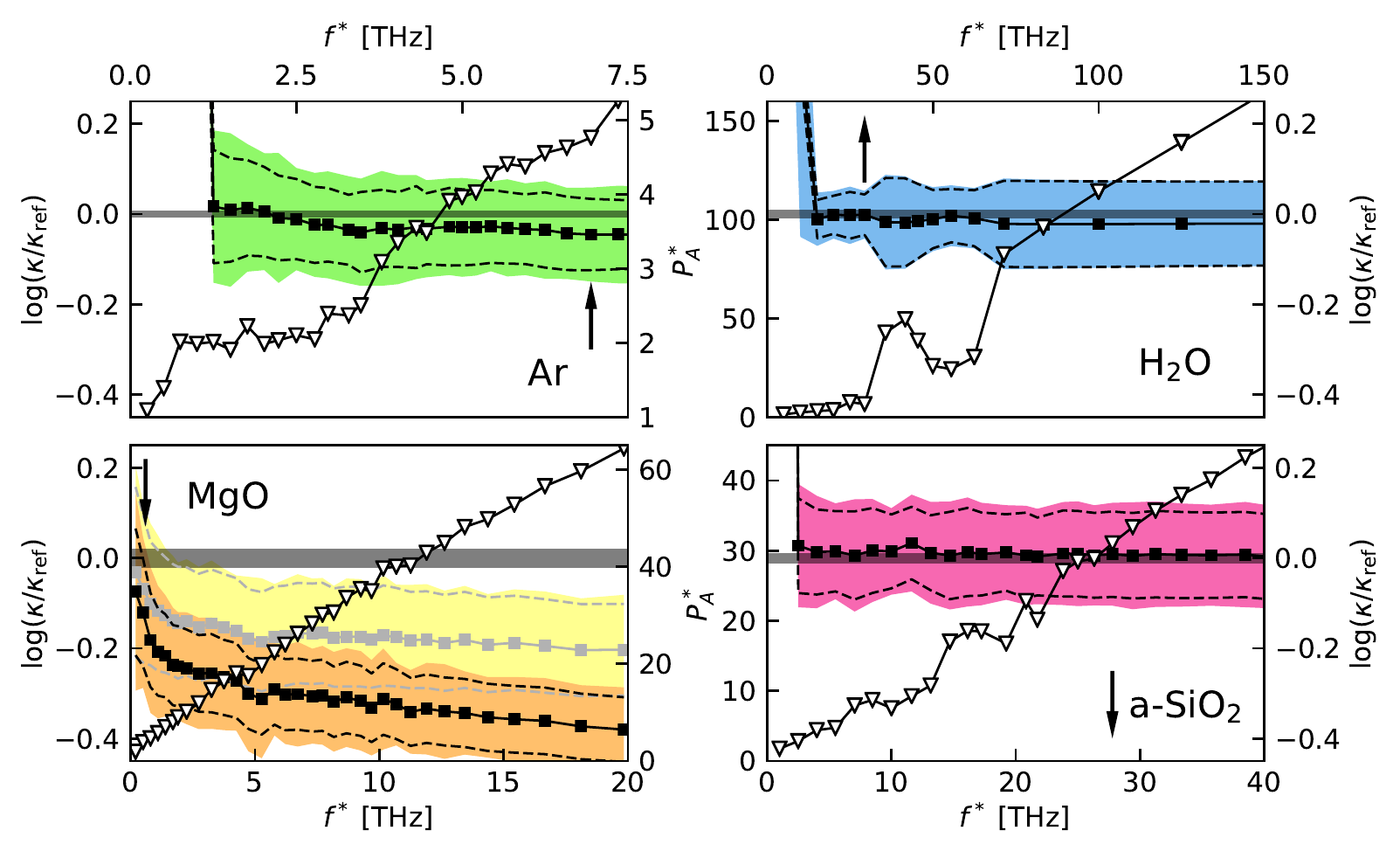}
\caption{Triangles: average optimal number of cepstral coefficients, $P_A^*$, as determined by the AIC, Eqs. \eqref{eq:P*} and \eqref{eq:AIC-P}, as a function of the cutoff frequency used for cepstral analysis, $f^*$ (see discussion just after Eq. \eqref{eq:AIC-P}). Squares: $\log(\kappa)$ resulting from a given choice of $f^*$ and of the corresponding value of $P_A^*$. All the values are averages performed over multiple $100\un{ps}$ long segments ($500\un{ps}$ for MgO) extracted from a $50\un{ns}$ long MD trajectory, as discussed in the text. The colored bands indicate the sample standard deviation and the dashed lines that resulting from our theoretical analysis (see Eq. \eqref{eq:sigma*}). The vertical arrows indicate the cutoff frequencies, $f^*$, used for the cepstral analysis in this paper (see Fig. \ref{fig:periodograms} and text). In the case of MgO, the data indicated with lighter colors are obtained using a number of cepstral coefficients twice as large as that provided by the AIC, $P^*=2P_A^*$. The data are referred to $\kappa_{\mathrm{ref}}$, which is the value of thermal conductivity obtained from direct integration of the current autocorrelation function in Eq. \eqref{eq:Green-Kubo}, combined with standard block analysis over the $50\un{ns}$ trajectory, and represented by the horizontal gray bands. Remember that the absolute error on $\log(\kappa)$ is the relative error on $\kappa$.
}
\label{fig:kappa_vs_fstar}
\end{figure}

Finally, in Fig. \ref{fig:kappa_vs_fstar} we report the dependence of the optimal number of cepstral coefficients, $P_A^*$, as a function of the cutoff frequency, $f^*$, along with the dependence of the resulting estimate of $\log(\kappa/\kappa_{\mathrm{ref}})$. $P_A^*$ increases with $f^*$. Notwithstanding, the estimated value of the heat conductivity, as well as its variance, is fairly insensitive on the precise value of $f^*$ as long as the latter is large enough as to encompass the lowermost prominent feature of the spectrum. Some more comments are in order for MgO. In this case the high thermal conductivity, due to the strong harmonic character of slowly decaying phonon modes, manifests in the form of a narrow peak centered at $f=0$, followed by a broad plateau that carries little spectral weight. This feature determines a more pronounced increase in the number of significant cepstral coefficients as $f^*$ increases and a corresponding increase of the bias when keeping $P^*$ at the value given by the AIC. The AIC is a less reliable indicator of the number of cepstral coefficients necessary to keep the bias low, in this case. By increasing this number by a factor of two or more, the bias decreases, as indicated by the results reported in lighter colors in Fig. \ref{fig:kappa_vs_fstar}, and eventually vanishes, as shown in Fig. \ref{fig:L-vs-P}.

\section{Conclusions and perspectives} \label{sec:conclusions}
We believe that the methodology introduced in this paper will finally open the way to the quantum simulation of heat transport, which has been considered out of its scope until very recently, particularly for strongly anharmonic and/or disordered systems. Its implementation is straightforward and its use robust, as the only parameter to be determined is the optimal number of cepstral coefficients, using \emph{e.g.} the Akaike's information criterion. Our methodology performs best when a low conductivity results from large cancellations in the integral of a highly oscillatory time auto-correlation function, such as in molecular liquids or amorphous solids, where simulation times of the order of $100\un{ps}$ seem sufficient to obtain accuracies of the order of $10\%$ in the estimated thermal conductivities. The performance is less spectacular in periodic crystals, where slowly-decaying strongly-harmonic phonon modes require longer simulation times and the ensuing sharp peak in the low-frequency region of the power spectrum requires a larger number of cepstral coefficients than predicted by the optimization of the AIC. Even so, simulation times of the order of a few hundred picoseconds seem sufficient to achieve a comparable accuracy. In the latter case, it is possible that a combination of the methodology introduced here with specialized techniques based on normal-mode analysis, such as that presented in Ref. \citenum{Carbogno2017}, will result in further improvements. Replacing the optimization of the AIC with more sophisticated and possibly more efficient approaches, such as \emph{e.g.} weighted multi-model inference techniques,\cite{Burnham2004,Claeskens2008} may also assist in this and other difficult cases. Finally, we expect that our methodology will impact on the simulation of any transport phenomena to which the Green-Kubo theory applies, such as ionic conduction, viscosity, and many others.

\section*{Acknowledgments}
This work was partially supported by the European Union through the \textsc{MaX} Centre of Excellence (Grant No. 676598). We are grateful to Dario Alf\`e and Giovanni Ciccotti for a critical reading of the first version of the manuscript. LE is grateful to Nicola Marzari for hospitality at MARVEL-EPFL, where this work took its final shape.

\section*{Author contributions statement}
LE contributed to the development of the theory, ran most of the simulations, wrote the data-analysis software, performed the data analysis, and contributed to write the manuscript. AM contributed to the development of the theory, ran some of the simulations, and contributed to write the manuscript. SB developed the theory, supervised the project, and wrote the manuscript.

\section*{Additional information}
\textbf{Competing financial interests:}
The authors declare no competing financial interests.

\bibliography{WFCbiblio}
\end{document}